\documentclass[english]{IEEEtran}
\usepackage[T1]{fontenc}
\usepackage[latin9]{inputenc}
\usepackage{xcolor}
\usepackage{rotfloat}
\usepackage{algorithm2e}
\usepackage{amsmath}
\usepackage{amssymb}
\usepackage{graphicx}
\usepackage{multicol}
\usepackage{soul}
\usepackage{url}
\usepackage{subcaption}
\usepackage{booktabs}

\makeatletter

\usepackage{babel}
\begin{document}
\title{Nested Multiple Instance Learning in Modelling of HTTP network traffic}
\author{Tom\'{a}\v{s} Pevn\'{y}~\IEEEmembership{Member,~IEEE,} and Marek D\v{e}di\v{c}
\thanks{Tom\'{a}\v{s} Pevn\'{y} was with the Department of Computer Science, Faculty of Electronics Engineering, Czech Technical University in Prague, 
    email: pevnytom@fel.cvut.cz}
\thanks{Marek D\v{e}di\v{c} was with the Department of Mathematics, Faculty of Nuclear Sciences and Physical Engineering, Czech Technical University in Prague and also with Cisco systems, Inc.
    email: marek995@seznam.cz}
}

\maketitle
\begin{abstract}
	In many interesting cases, the application of machine learning is hindered by data having a complicated structure stimulated by a structured file-formats like JSONs, XMLs, or ProtoBuffers, which is non-trivial to convert to a vector / matrix. Moreover, since the structure frequently carries a semantic meaning, reflecting it in the machine learning model should improve the accuracy but more importantly it facilitates the explanation of decisions and the model. This paper demonstrates on the identification of infected computers in the computer network from their HTTP traffic, how to achieve this reflection using recent progress in multiple-instance learning. The proposed model is compared to complementary approaches from the prior art, the first relying on human-designed features and the second on automatically learned features through convolution neural networks. In a challenging scenario measuring accuracy only on unseen domains/malware families, the proposed model is superior to the prior art while providing a valuable feedback to the security researchers. We believe that the proposed framework will found applications elsewhere even beyond the field of security.
\end{abstract}

\IEEEpeerreviewmaketitle

\section{Motivation}
This paper presents a method to identify infected computers from their HTTP traffic, which can be collected on the network perimeter, in the cloud proxy, or by an antivirus installed on monitored computers. Although the model assumes visibility of URL strings, which is decreasing due to the popularity of HTTPs, we believe that the problem is still important and interesting because: (i) the solution presented here is general and the framework can be used for other problems with a hierarchical structure; (ii) there is an established prior art which is well developed and non-trivial to outperform; (iii) URLs can be still collected directly at the endpoint before being encrypted, for example by an antivirus engine, virtual private network agents, by browser extensions, or by a white-hat version of a man in the middle attack employed by some companies.

The prior art on using machine learning (ML) on URL strings is vast and relevant works are briefed in Section~\ref{sec:related}. URL strings are challenging for ML models since they have an internal structure consisting of several blocks, consisting of a variable number of tokens including zero.  Coping with this variability is a challenge solved in the most prior art by two orthogonal approaches: the first~\cite{kruegel2003anomaly,zarras2014automated,machlica2017learning,li2018method} converts the URL to a Euclidean space $\mathbb{R}^d$ using a set of human-designed features, which enables to use off-the-shelf machine learning systems such as Random Forests or Support Vector Machines. The second~\cite{saxe2017expose} avoids human-designed features by converting the URL string to a matrix with one-hot encoded characters.\footnote{In this representation each column corresponds to one character from the URL with the value one on the row with an index of the character and zero elsewhere.}  Although this allows using convolution neural networks (CNN) or recurrent neural networks, they have to learn how to parse and interpret the structure of the string, which makes the learning unnecessarily complicated and opaque since the structure is known.

This work tries to fix the shortcomings of the prior art and extend it along  three directions.
\begin{enumerate}
	\item It presents a new model of URL strings, which (i) avoids human design features like CNN, (ii) removes the need to truncate or pad URL strings to fit the predefined number of characters, and (iii) exploits the structure of URL strings.
    \item Unlike the most prior art modeling just individual URL strings, here the model utilizes complete HTTP traffic of a single computer. This gives it the ability to detect infections that changes the distribution of traffic on otherwise legitimate servers (e.g. adware) (which models scrutinizing URL strings independently cannot do).
    \item The hierarchy of data reflected in the model allows to extract indicators of compromise and explain the decision by identifying parts of a sample responsible for positive classification. 
\end{enumerate}
The model is learned end-to-end from labels on the level of a computer. The motivation for this is manyfold: (i) according to Ker's laws~\cite{ker2006batch,ker2017thesquare} it should improve accuracy of classification, which is based on more data; (ii) it moves the granularity of labeling from individual URL strings to the whole computer, which is simpler and more accurate;\footnote{Imagine that you know the computer is infected and you should label an HTTP request to \url{google.com}. This request can be due to search invoked by a user or by malware checking connection to the internet.} (iii) using multiple URL strings gives the classifier an ability to detect malware, that communicates only with legitimate servers but changes a distribution on them (see~\cite{pevny2019approximation} for a formal proof). This type of malware might be undetectable by classifiers using only single URLs, as they look perfectly normal.

The proposed model is experimentally compared to representative work of prior art based on (i) hand-designed features and random forests~\cite{machlica2017learning} (further called  R. Forest model) and (ii) convolution neural network~\cite{saxe2017expose} (further called by its name eXpose). Experiments are carried on public and private datasets and they are designed to evaluate accuracy on future and unseen malware samples (Grill test~\cite{grill2016learning}), accuracy in identifying the type of infection, and dependency on the number of labeled samples. 

The paper is structured as follows. Section~\ref{sec:MIL} reviews multi-instance learning and particularly the method~\cite{pevny2017using} on which the proposed structured model is based. Section~\ref{sec:model} describes the proposed structured model of URL requests and the  traffic of computers. Implementation of the model and prior art, datasets, and other experimental details are summarized in Section~\ref{sec:Details}. Experimental results are presented in Section~\ref{sec:experimental} and Section~\ref{sec:conclusion} concludes the work.

\section{Multi-instance learning}

\label{sec:MIL}
\begin{figure}
	\centering
	\includegraphics[width=0.9\columnwidth]{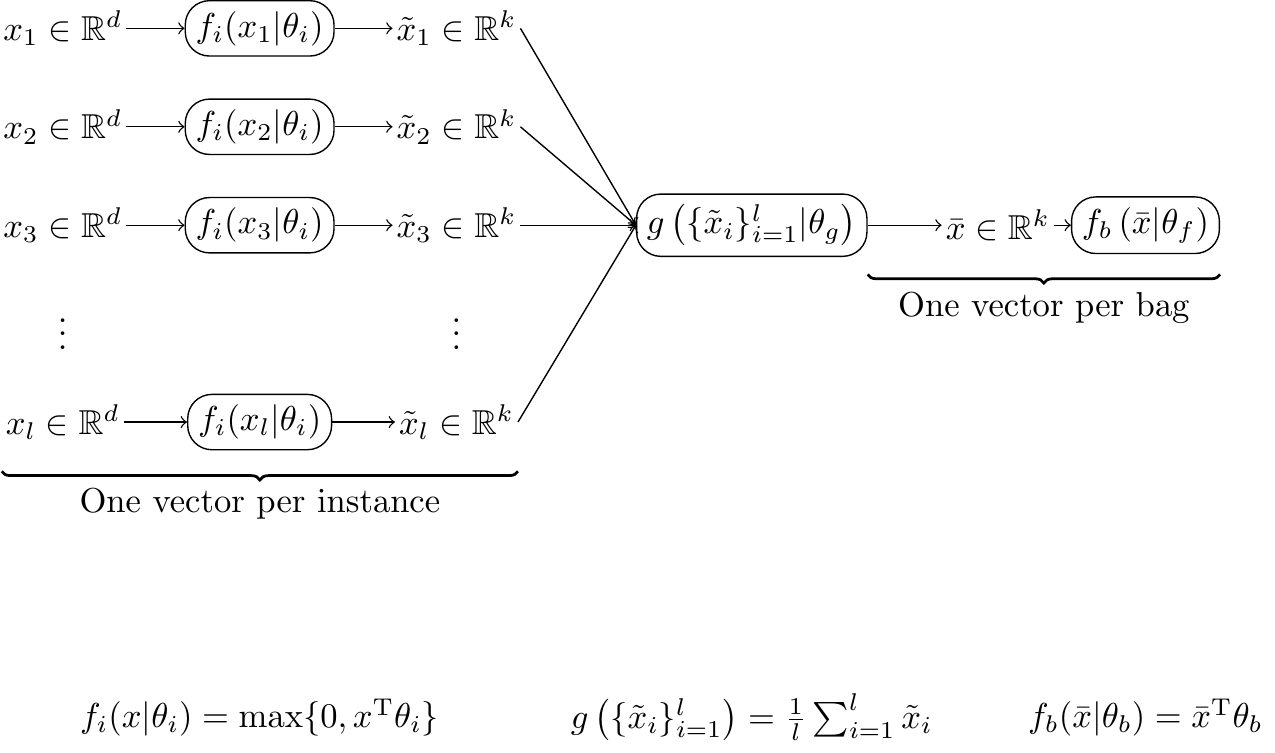}
	\caption{\label{fig:mil-outline}Outline of a classifier for multiple-instance problems proposed in~\cite{pevny2017using,edwards2017towards}.}
\end{figure}
The problem of \emph{multi-instance learning} (MIL) has been introduced
in~\cite{dietterich1997solving} to solve the problem where a sample
(in MIL nomenclature called \emph{bag}) was composed by a set of vectors
(in MIL nomenclature called \emph{instances}) of an arbitrary number,
but of a fixed size (dimension). In pioneering work~\cite{dietterich1997solving} it has been assumed
that there exist labels on individual instances, but during training only labels on the level of bags are known. Later works adopt a more general formulation of the problem by~\cite{muandet2012learning}, which assumes the sample
to be a probability distribution observed through a finite number of realizations
of a corresponding random variable. The very same work has proposed to solve this problem using a combination of support vector machines with a probabilistic kernel~\cite{christmann1020universal}.
While their classifier converges to the optimal detector, it does not scale, as its worst-case complexity is $O(l^{3}b^{2}),$ where $l$ is the number of samples in the training set and $b$ is the average size of the bag.

Refs.~\cite{pevny2017using,edwards2017towards} have independently proposed a simple classifier utilized in this work and outlined in Figure~\ref{fig:mil-outline}, which according to extensive experiments outperforms most prior art~\cite{pevny2017using}. Denoting a single sample (bag) as $\boldsymbol{b}=\{x_{1},\ldots,x_{b}\},$ $x_{i}\in\mathbb{R}^{d}$, the classifier is implemented by two feed-forward neural networks $f_{i}$ and $f_{b}$ with an element-wise aggregation $g$ between them. The first network $f_{i}$ embeds each instance $x_{i}$ into a $k$-dimensional space as $f_{i}(x_{i})=\tilde{x}_{i}\in\mathbb{R}^{k}$, then the element-wise aggregation combines all projected instances $\{\tilde{x}_{1},\ldots,\tilde{x}_{l}\}$ into a single vector $\overline{x}\in\mathbb{R}^{k}$ of the same dimension $k$, which means the whole bag of an arbitrary number of instances is represented by a single vector $\overline{x}\in\mathbb{R}^{k}$. Finally, the network $f_{b}$ provides the final decision.

The simplest implementation of this classifier uses  single non-linear layer with rectified linear units for $f_{i},$ an element-wise average for aggregation $g,$ and finally a linear function for $f_{b}.$ The final classifier can be written as
\begin{equation}
    f(\boldsymbol{b})=\frac{1}{l}\left(\sum_{i=1}^{b}\mathrm{relu}(\theta_{i1}^{\mathrm{T}}x_{i}),\ldots,\sum_{i=1}^{b}\mathrm{relu}(\theta_{ik}^{\mathrm{T}}x_{i})\right)\theta_{b.}
\end{equation}
Its main advantage with respect to prior art~\cite{amores2013multiple} is that since the whole scheme is differentiable, all parameters including those of the instance-projection function $f_{i}$ can be optimized using stochastic gradient descend and its variants. Furthermore, it has been shown in~\cite{pevny2019approximation} that the construction is dense in the space of continuous functions from probability measures over $\mathbb{R}^{d}$ to $\mathbb{R}$.

\section{Hierarchical model of a traffic of a computer}
\label{sec:model}
\begin{figure*}[t]
    \centering
    \includegraphics[width=0.95\textwidth]{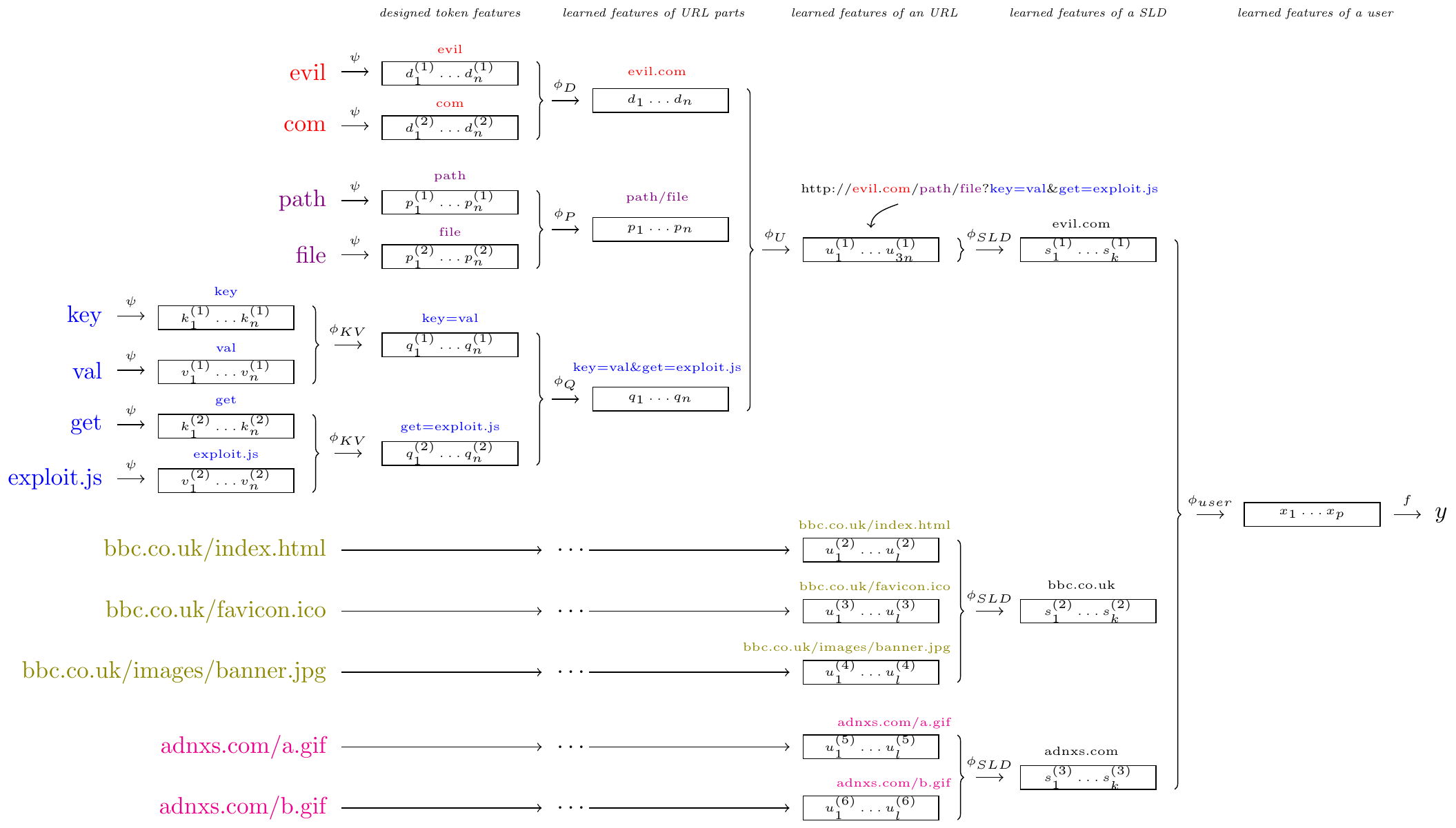}
    \caption{\label{fig:computer}Hierarchical model of the network traffic of a single computer. The computer is modeled by a set of remote servers and each remote server is modeled by a set of messages (connections) from the given computer to it.}
\end{figure*}

This section describes an application of the multi-instance learning approach to identify an infected computer based on its HTTP network traffic observed on a network perimeter. The explanation starts by modeling the traffic under the assumption that each connection (message) is already represented by a vector. Then, a model of URL is presented under the assumption that each token within is represented by a vector. Finally, a representation of a token is detailed. Putting all three pieces together in reverse order gives rise to the full model demonstrated in this work. It also shows the power and flexibility of the modeling framework proposed in~\cite{pevny2016discriminative}.

\subsection{Embedding the network traffic of computers}
\label{sub:Traffic} The model of computers' traffic is adopted from our previous work~\cite{pevny2016discriminative}. It assumes that the network traffic of a single computer can be modeled by a set of remote servers (identified by a domain) with which the computer has communicated with, i.e. to which it has issued a request over HTTP protocol. In an example in Figure~\ref{fig:computer}, the computer is modelled by communication to three domains \texttt{evil.com}, \texttt{bbc.co.uk}, and \texttt{adnxs.com}. Similarly, each hostname is modeled by a set of requests a particular computer has issued to the server. Again, in Figure~\ref{fig:computer} the domain \texttt{bbc.co.uk} is modelled by three HTTP requests \texttt{bbc.co.uk/index.html}, \texttt{bbc.co.uk/favicon.ico}, and \texttt{bbc.co.uk/banner.jpg}.

The proposed model simplifies the complexity of the network traffic along two axes: (i) it assumes that domains do not interact with each other, i.e. it cannot properly model mash-up sites; (ii) it does not take the time continuity in the account, which means that it cannot model time dependencies within sequences. The first simplification is mainly technical, as it is difficult to attribute an HTTP request in mash-up to the originator. The second removes problems with catastrophic forgetting and gradient explosion/diminishing in recurrent neural networks.

Despite these simplifications below experiments demonstrate that the expressive power remains high to identify almost all infected computers. We hypothesize, that it is because the infected computer has either a very distinct probability distribution of requests issued by malware (for example URLs sent to a command and control server has a special format) or the infection changes the distribution of types of servers with which the computer interacts (for example adware downloads a large number of ads albeit from legitimate services).

\subsection{Embedding URL strings}
The above model of computer traffic assumes that URL strings are already embedded in a Euclidean space. Since this is not straightforward, this section fills this gap by showing how this embedding is implemented (and optimized) using MIL formalism.

The URL string is viewed as a Cartesian product (concatenation) of models of hostname, path, and filename, and query parameters respectively (note that path and query parts are optional).\footnote{The  scheme, port, and users are not represented, but they can be added if needed.} Since each part can contain an arbitrary number of tokens (parts of hostname, file path, key-value pairs), the MIL framework is used again to handle this variability. Finally, key-value pairs in the query part are represented as a Cartesian product (concatenation) of representations of a key token and a value token.

In URL string 
\begin{center}
{\smaller \texttt{http://evil.com/\allowbreak path/\allowbreak file?\allowbreak key=val\&\allowbreak get=exploit.js}}\newline
\end{center}
used in Figure~\ref{fig:computer}, the hostname \texttt{evil.com} consists of tokens \texttt{evil}, \texttt{com} (modeled by the first MIL problem); path and filename consist of tokens \texttt{path} and \texttt{file} (modeled by the second MIL problem); and finally query parameters consist of pairs \texttt{key=val} and \texttt{get=explit.js} (modeled by the third MIL problem). Pair \texttt{get=explit.js} is represented by a key token \texttt{get} (one model) and value token \texttt{explit.js} (another model). What remains to explain is how to embed individual string tokens to Euclidean space, which is left to the next section.

The proposed representation of URL strings and the network traffic offers a lot of flexibility, as a representation of each bag on each level of the hierarchy is parametrized by a feed-forward neural network. Specifically (see Figure~\ref{fig:computer}), functions $\phi_{D},$ $\phi_{P},$ $\phi_{K},$ and $\phi_{V}$ embed tokens of hostname, path, and keys and values in query respectively. Similarly functions $\phi_{Q},$ $\phi_{SLD},$ and $\phi_{user}$ embeds higher-level concepts such as key-value pairs in the query, URL strings, all traffic to a domain, all domains the computer has contacted, and finally, $f$ utilizes the topmost embedding and classifies the computer. All feed-forward neural networks $\phi_{\cdot}$ can contain arbitrary number of layers and can implement recently proposed extensions such as dropout~\cite{srivastava2014dropout}, layer normalization~\cite{ba2016layer}, residual connections~\cite{he2016deep}, etc. Needless to say that in the below experiments, none of these extensions have been used and all functions $\phi_{\cdot}$ were implemented as a single layer with 80 neurons and a relu nonlinearity.

\subsection{Embedding string tokens}
What remains to explain is how to embed individual string tokens to Euclidean space. Although this is a subject of extensive research in natural language processing and language translation, not all solutions can be trivially adapted to the domain of URL modeling, as the language of tokens in URLs is several orders of magnitudes larger because a token can be an arbitrary sequence of allowed characters including Unicode. Since this problem is outside of the scope of this work, the experimental section uses a very simple representation. Each token is represented as a histogram of indexes of trigrams. Since the number of trigrams can be very high (in theory $256^{3}$ as Unicode is converted to its byte representation) the dimension is decreased by applying the modulo operation. In the experimental section, the modulo is $2053$. The rationale behind this number is that (i) it is a prime, which is important for the distribution in hashing and (ii) the number seems to be a good trade-off between the complexity of the network and accuracy of the solution.

We have experimented with an alternative representation inspired by eXpose~\cite{saxe2017expose} , where a token is seen as a sequence of one-hot encoded characters. This sequence is embedded into Euclidean space by first applying the convolution windows of size three and then reducing its output by elementwise average and maximum. Despite the complexity of this solution is  4-5 times higher than that of based on trigrams, and it has required the implementation of a custom convolution operator that can handle tokens of different sizes, in our setting it was inferior to an embedding based on trigrams. It might be possible to improve this approach with sufficient tweaking, but we leave this option to future work.

\subsection{Extracting indicators of compromise}
\label{subsec:IOC}
If we assume that the HTTP requests caused by malware are added to those caused by user interaction, then infected computers emit a mixture of normal and malicious requests. This means that neurons inside the trained network should be sensitive to malware tokens / URL strings / domains and insensitive to those of clean computers. This means that each neuron is a weak indicator of compromise (IOC). By inspecting the tokens / URL strings / domains to which neurons on different level of hierarchies are sensitive have several benefits: (i) it helps to understand and verify the function of neural network; (ii) it can reveal some phenomenons security analysts were not aware of; and finally (iii) it can help explain the decision and provide context, especially if the type of traffic can be described in a human understandable language. Notice that the hierarchical structure of the proposed model allows to select the level of granularity (tokens / URL strings/domains) a security researcher is interested in.

The exact algorithm extracting important parts of the traffic at a particular level of detail works as follows. 
\begin{enumerate}
    \item Calculate the average output of neurons (at that given level) on a normal traffic, further denoted by $o_{\textrm{normal}}.$
    \item Calculate the output of same neurons on the traffic of infected computers, further denoted as $\{o^i_{\textrm{infect}}\}_{i=1}^l.$ \footnote{Note that while $o_{\textrm{normal}}$ is a single vector with dimension equal to the number of neurons, $\{o^i_{\textrm{infect}}\}_{i=1}^l$ is a set with a cardinality equal to the number of URL strings / tokens or domains.}
    \item For $j^{\mathrm{th}}$ neuron (dimension of vectors $o_{\cdot}$), calculate the normalized score as $\frac{o^i_{\textrm{normal},j}}{o_{\textrm{infect},j}}.$ Then the URL strings / tokens or domains with score much greater or much smaller than one are the most characteristic for a given type of infection. Note that scores around one are non-indicative, as they are similar to the traffic of normal computers.
\end{enumerate}

The above algorithm assumes that the malware adds but does not remove HTTP requests. We can imagine cases, where such removal occurs, for example, if malware disables an anti-virus engine, which then ceases updating itself. But missing these updates are not indicative of infection since not all users have an anti-virus installed. We, therefore, believe that the assumption on additivity does not have a significant impact on the extraction of IOCs.

\subsection{Explaining the decision}
\label{subsec:explanation} Neural networks have a bad reputation of being a black-box model without any possibility to extract any explanation of their decision. In intrusion detection, this feature might prevent widespread adoption, as people have a tendency not to trust in systems that they do not understand. Moreover, providing an explanation of the security incident to the analyst might simplify and speed-up their investigation.

Similarly to the extraction of IOCs, the explanation exploits the structure of the model and it also relies on the assumption that malware can only add HTTP requests, but it cannot remove them. \footnote{The question underlining this assumption is whether interfering with user's HTTP requests can make the malware visible to the user, since the user might notice that the device works differently, which might trigger his investigation leading to the identification of infection.} This has the implication that by removing malware HTTP requests the decision of a neural network can be changed from infected to clean. Although finding the smallest number of such requests is likely an NP-complete problem, a greedy approximation inspired by~\cite{silveira2010urca} performs surprisingly well.

The explanation is iterative, where in each iteration a set of requests to the same hostname causing the biggest decrease of the classifier's output are removed (in our implementation positive means infected). The algorithm stops when there are no further requests. The set of all removed requests is grouped by their hostname and the returned explanation will have the form: ``This computer was found infected because it has communicated with these servers''. This explanation can be further augmented by annotating the type of traffic to which neurons are sensitive to as described in the previous subsection, and use the activity of these neurons to improve the description.\footnote{The explanation on the level of requests to servers has been shown in our prior work in~\cite{pevny2016discriminative}. Due to privacy concerns, this work demonstrates this on a publicly available CSIC~\cite{perez2010applying} dataset on a lower level of tokens in URL.}

\section{Implementation and experimental details}
\setlength{\tabcolsep}{5pt}
\begin{table*}[ht]
	\centering
	\begin{tabular}{lrrrrrrrrrr}
		\toprule
		                       & \multicolumn{5}{c}{Training set}                        & \multicolumn{5}{c}{Testing set} \\
		\cmidrule(lr){2-6} \cmidrule(l){7-11}
		                       & \#camp. & \#users & \#windows & \#domains & \#URLs     & \#camp. & \#users & \#windows & \#domains & \#URLs \\
		\midrule
		2  trojan              & 3       & 172     & 1562      & 6         & 6644       & 3       & 17      & 47        & 4         & 62 \\
		3  ad injector         & 46      & 15196   & 566709    & 451       & 9262340    & 33      & 1804    & 11684     & 214       & 89593 \\
		5  ransomware          & 1       & 15      & 181       & 1         & 270        & 1       & 4       & 5         & 1         & 7 \\
		6  malicious           & 18      & 14725   & 54533     & 594       & 422362     & 14      & 1442    & 3276      & 89        & 33773 \\
		8  PUA                 & 21      & 18633   & 413314    & 189       & 5549824    & 14      & 2449    & 22098     & 140       & 251955 \\
		9  malware             & 3       & 35      & 10036     & 28        & 30363      & 2       & 4       & 405       & 12        & 1516 \\
		11 information stealer & 13      & 5614    & 154761    & 175       & 1154500    & 7       & 526     & 7240      & 19        & 65701 \\
		12 mal. cont. dist.    & 18      & 8962    & 35993     & 716       & 121558     & 15      & 646     & 1379      & 138       & 6559 \\
		13 scareware           & 5       & 865     & 72784     & 12        & 77559      & 3       & 123     & 3646      & 7         & 3824 \\
		14 money scam          & 2       & 494     & 947       & 74        & 1220       & 2       & 23      & 42        & 20        & 53 \\
		15 anon. software      & 2       & 550     & 19368     & 69        & 31089      & 1       & 84      & 926       & 53        & 1716 \\
		16 banking trojan      & 2       & 6       & 5106      & 12        & 10578      & 1       & 2       & 294       & 10        & 620 \\
		17 spam tracking       & 8       & 1530    & 2860      & 305       & 20889      & 7       & 108     & 143       & 64        & 598 \\
		18 click fraud         & 8       & 584     & 3882      & 60        & 10344      & 5       & 34      & 186       & 23        & 385 \\
		19 cryptominer         & 1       & 9       & 18213     & 9         & 194173     & 1       & 3       & 524       & 9         & 6475 \\
		\midrule
		legitimate             & ---     & 4450240 & 261823324 & 19306864  & 3597464926 & ---     & 1494471 & 13626770  & 2110986   & 177786802 \\
		\bottomrule
	\end{tabular}
	\caption{\label{tab:trafficsummary}Summary of infected computers and the infection types in the training set. "\#camp." corresponds to the number of families, "\#users" to the number of computers, "\#windows" to the number of 5-minute windows of traffic of a single computer associated with the infection, "\#domains" to the number of network domains and "\#URLs" to the number of URLs associated with a virus. Abbreviation "mal. cont. dist." stands for malicious content distribution and "anon. software" stands for anonymization software.}
\end{table*}

\label{sec:Details}
This section details: the implementation of the presented models together with the selected prior art; the datasets used for evaluation; grill-test used to estimate accuracy on unseen malware; and finally the metric used to compare the classifiers. Unless said otherwise, each experiment has been repeated ten times.

\subsection{The hierarchical model}
The proposed hierarchical model based on multi-instance learning was implemented in the Julia language~\cite{Julia} using the Flux.jl~\cite{innes:2018} library for automatic differentiation and the authors' library supporting nested multiple instance learning models and their Cartesian products available at \url{https://github.com/pevnak/Mill.jl}. The experimental section compares two models based on the proposed MIL framework.

\emph{MIL-5min} is the classifier advocated in this work. It classifies a computer based on all network traffic observed during 5-minute windows, as has been described in the previous section and as is outlined in Figure~\ref{fig:computer}. Its main advantage is that (i) it can be trained on coarse labels (whole computer vs. individual URLs) and (ii) it has more information (multiple URLs) upon which it can base its decision.

\emph{MIL-URL} is a submodule of the MIL-5min model providing the decision on individual URL strings. It is outlined in the upper part of Figure~\ref{fig:computer} showing the representation of URL strings. The rationale behind introducing this model is that it allows direct comparison to the prior art, which works mostly on individual URL strings. The main drawback of this model is that it requires labels on the level of individual URL strings, similarly to most of the state of the art. The comparison of its accuracy to MIL-5min also shows that the drop in accuracy by training on coarse labels is negligible.

Unless said otherwise, individual feed-forward networks, $\phi_{\cdot},$ consisted of a single layer of 80 neurons with ReLU non-linearity. The aggregation of bags used mean and maximum simultaneously, therefore increasing the dimensionality from 80 to 160. The rationale behind this was that both the mean and the maximum aggregation functions have their advantages. Mean is very good for the case when malware is abusing legitimate services and infected computer exhibits change in the probability distribution of contacted types of servers~\cite{pevny2019approximation}. Maximum is good when malware contacts single or few servers with a very distinct pattern, for example, a command and control channel. Utilizing both aggregation functions simultaneously provides the best of both worlds.

The last layer of the neural network is augmented by a linear layer with 16 neurons, as computers were classified into 16 classes according to the type of malware and one class deeming the computer clean. The loss function was the usual cross-entropy, with the only exception being that the error on clean computers had a weight $1 - w_{\textrm{pos}}$, while the error on infected computers had a weight of $w_{\textrm{pos}}.$ The rationale behind this is that in real use, the detection system will observe two to three orders of magnitude higher number of clean computers than infected ones. A high false-positive rate is therefore devastating as the network operator would be flooded with false alarms. Unless said otherwise, $w_{\textrm{pos}} = 0.01$.

The ADAM~\cite{kingma2014adam} stochastic gradient descent method with default settings was used with batch size $2\times 1000$ samples when the sample was a computer (MIL-5min model) and $2\times 5000$ samples when the sample was a single URL string (MIL-URL model). The gradient descend was run for 50 000 iterations. Because loading the data was very time consuming, the stochastic gradient descends used a circular buffer of size 5, which means that every minibatch was reused 5 times. This means that although the SGD used 50 000 steps, it has seen only 10 000 "new" mini-batches.


\subsection{Prior art}
The proposed solution has been compared to two approaches --- manually designed features used in a random forest classifier~\cite{machlica2017learning} (called R. Forest) and an approach based on convolution neural network~\cite{saxe2017expose} (called eXpose). They were selected as they represent state complementary approaches in the prior art.

R. Forest classifier~\cite{machlica2017learning} uses a set of 398 hand-designed features that are used with a random forest classifier to separate URL strings of benign and malicious applications. This approach is a good prototype of an industry workhorse, as random forests are very robust and hand-designed features allow to incorporate a lot of domain knowledge into the solution. The set of features proposed in~\cite{machlica2017learning} has been also used for example in~\cite{li2018method}. We have reimplemented these features by ourselves, but thanks to authors we could verify their correctness as authors have provided us with a set of URLs and corresponding feature values. Random Forests used the implementation from \url{https://github.com/bensadeghi/DecisionTree.jl} v$0.8.1$ in the Julia~\cite{Julia} language. In all experiments below, each forest contained 100 trees with a maximum depth 30 and we have left all settings to default, which authors of~\cite{machlica2017learning} confirmed as reasonable settings. To maximize the diversity of trees within the forest, each tree has been trained on its random subset of all training URLs. Although we were not able to train the forest on all URLs in the training set, as we have been limited by the 64Gb of memory of a single m4.x8 instance, the model still used $2.4\times$ more labeled samples than MIL-URL classifier. In all experiments, we have trained each on $1\,200\,000$ negative and the same number of positive samples. 

eXpose~\cite{saxe2017expose}, inspired by the success of convolution in digital images, builds a classifier of URLs using a convolution neural network. Specifically, eXpose truncates or pads all URL strings to 200 characters such that they all have the same size. Using a one-hot encoding of characters, a URL is then converted into a binary matrix of size $200 \times 256$, which allows for using a common stack of convolution, reduction, and fully connected layers used in image recognition. eXpose was implemented exactly as described in~\cite{saxe2017expose} with the only difference being that samples have been classified into 16 categories (clean + 15 malware categories) instead of benign/malicious. eXpose was implemented in the Julia~\cite{Julia} language in the Flux.jl~\cite{innes:2018} library for neural networks. We have used the ADAM optimizer~\cite{kingma2014adam} with a minibatch of size 256 (mandated by the limit of GPU memory) and it was allowed to train for $50\,000$ iterations on Amazon's p2.xlarge GPU instance.

\subsection{Corporate dataset}
The main dataset of HTTP traffic was collected from more than 500 large customers of Cisco's Cognitive Threat Analytics (CTA)~\cite{cta} during one month from $5\mathrm{th}$ October 2017 till $3\mathrm{rd}$ November 2017. The inherent limitation of the CTA engine is that it discards 93\% of the observed traffic and keeps only the most suspicious part, which is 7\%. Nevertheless, the dataset still contains more than $10^{13}$ URL strings. The training part contains all traffic collected in October and the testing set contains traffic collected in $3^{\textrm{rd}}$ November. 

URL strings were labeled using Cisco's internal blacklist based on hostnames combined with regular expressions. The blacklist is curated by senior security officers and it is accurate in the sense that it contains minimal false positives (connections made by legitimate applications but attributed to the malware), but we admit that there might be false negatives, i.e. it can contain URL strings made by malware yet classified as legitimate. We argue that almost all datasets will suffer from this type of error for two reasons. First, it is at this moment impossible to obtain an accurately labeled large dataset. Second, as was mentioned in the motivation of this paper, even senior officers can have difficulties in labelling some URL strings. A typical example is connections to \texttt{google.com} made by the malware to check if it is connected to the internet.

We have a preferred private blacklist over labeling using the public services such as Virus Total (VT), since (i) they are prone to false positives, (ii) we do not have sufficiently high quotas to ask VT for every observed domain, (iii) it is not trivial to infer labels from the results provided by Virus Total answers~\cite{sebastian2016avclass}, (iv) the labeling will suffer from the same problems as our internal blacklist. Again, we do not think that possible false negatives in the dataset make the results less credible. Details and statistics about the dataset are shown in Table~\ref{tab:trafficsummary}.
Although the testing set does not contain any new malware campaign, it contains approximately 12.57\% domains that are not present in the training set.

\subsection{HTTP dataset CSIC 2010}
The CSIC 2010 dataset~\cite{perez2010applying} is a public collection of URL strings created to test web attack protection systems. It contains automatically generated web requests targeted to an e-commerce web application developed by the Group of Information and Communication Technologies at the Institute for Physical and Information technologies.

The dataset contains 36 000 normal URL strings and more than 25 000 anomalous ones including samples of following attacks: SQL injection, buffer overflow, information gathering, files disclosure, CRLF injection, XSS, server-side include, parameter tampering, etc.

The dataset contains traffic only to a single domain from a single computer, and the number of types of attack is very limited. It, therefore, allows assessment of accuracy only on individual URLs and it, therefore, does not mimic well the scenario of interest of this work. On the other hand, it is accurately labeled and its publicity is good for reproducible science, which is difficult in the field of network security. The experiments on this dataset presented in Section~\ref{subsec:exp:csic} should be therefore treated as a supplement to main experiments on the Corporate dataset which contains nine orders of magnitude more URL strings.

\subsection{Grill tests}
In this paper, all classifiers have been always evaluated on future data to observe the effect of aging. Yet, since labels in future data are created using the same blacklist as data in the training set, they will be correlated as a large number of domains and malware families will be present both in the training and testing set. This makes it difficult to estimate how well the classifier detects new types of malware, infections, and migrations to new domains, which is precisely the type of accuracy practitioners are interested in.

An experimental protocol originally proposed in~\cite{grill2016learning} aims to rigorously measure this type of generalization. Below, we present its minor variation allowing a straightforward comparison to the baseline, where the labels in the training set are not manipulated. Bellow, the test is called \emph{Grill test} as a tribute to the main author and also because it puts the classifiers on the edge of their capabilities.

The Grill test was executed on two levels: hostname and malware families. For simplicity, it is explained for hostnames, but its variant with malware families is straightforward.

Grill test assumes that each positive (malicious) sample is attached to a hostname, which in the case of URLs is trivial. In the beginning, all hostnames of positive samples from the testing set are randomly divided into $k$ folds (in this paper $k=5$). Then, $k$ classifiers are trained, where the training set of each classifier has either \emph{relabelled} or \emph{removed} samples with a hostname matching those in the corresponding fold. During classification, if a true label of a sample is positive, its hostname has to belong to one of the folds\footnote{Note that this property is ensured by the fact that folds are created from data in the testing set} and it is classified using the classifier which did not have this hostname in its training set. If a sample is negative, its hostname does not belong to any fold and the output is calculated as an average of all $k$ classifiers. Thus although in the Grill test a total of $k$ classifiers are trained, they act as a single classifier and positive samples are always classified using the classifier with corresponding hostnames removed or relabelled in its training set.

The rationale behind relabelling and removing is the following. \emph{Relabelling} simulates a scenario where the blacklist is not accurate and it contains false negatives, which means that the training set contains malware samples labeled as benign. \emph{Removing} simulates a scenario, where a new malware family appears after the classifier was trained, i.e. samples are not present in the training set at all.

We believe that the Grill test is important for practitioners, as it demonstrates how the classifier can detect new threats. Moreover, since the protocol used in this paper preserves the number of positive and negative samples in testing, it can be directly compared to the case, where training and testing data are shifted just by time. 

The Grill test used a Corporate dataset with training data collected in October and testing data in $3^{\textrm{rd}}$ November.

\subsection{Evaluation metrics}
Although all classifiers are trained to solve multi-class problems, they are firstly compared to a binary problem of identifying clean vs. infected computers. This is because, in reality, the false alarm / missed detection is more important than incorrect identification of malware. On this binary problem, classifiers are compared using the Precision-Recall curve~\cite{rijsbergen1975} known from information retrieval. The rationale behind this is that unlike the ROC curve, it is sensitive to class ratio and it shows how many infected computers can be identified (recall) and what is the fraction of truly infected computers out of the total number of computers classified as infected (precision). Since most experiments are repeated ten times, the plot contains average PR curves together with an area of $\pm 1$ standard deviation.

Precision-Recall curves (PR curves) are estimated and plotted on two levels of granularity. The \emph{microscopic} operates on the level of individual URL strings, and its main purpose is to observe how well are classifiers trained. However, for practical purposes the \emph{macroscopic} precision-recall curve on the level of users is more important, where all traffic of a single user is treated as a single sample. This shows how many computers would require attention by a diligent staff investigating every security incident. 
The decisions of classifiers on individual URLs (MIL-URL, Conv. Random Forest) or on individual five-minute windows (MIL-5min) are aggregated using maximum, i.e.
\begin{equation*}
    \max_{i \in \mathcal{I}} f(x_i),
\end{equation*}
where $f(x)$ is the output of the classifier on a sample $x$ (URL or five-minute window). An alternative to maximum aggregation is mean~\cite{ker2006batch}, which would correspond to counting alarms, or an aggregation learned from the data~\cite{pevny2015optimizing}. But according to discussions with security officers maximum well mimics functionality of real intrusion detection systems triggering on every alarm.

\begin{figure*}[ht]
    \centering
    \includegraphics[width=0.8\textwidth]{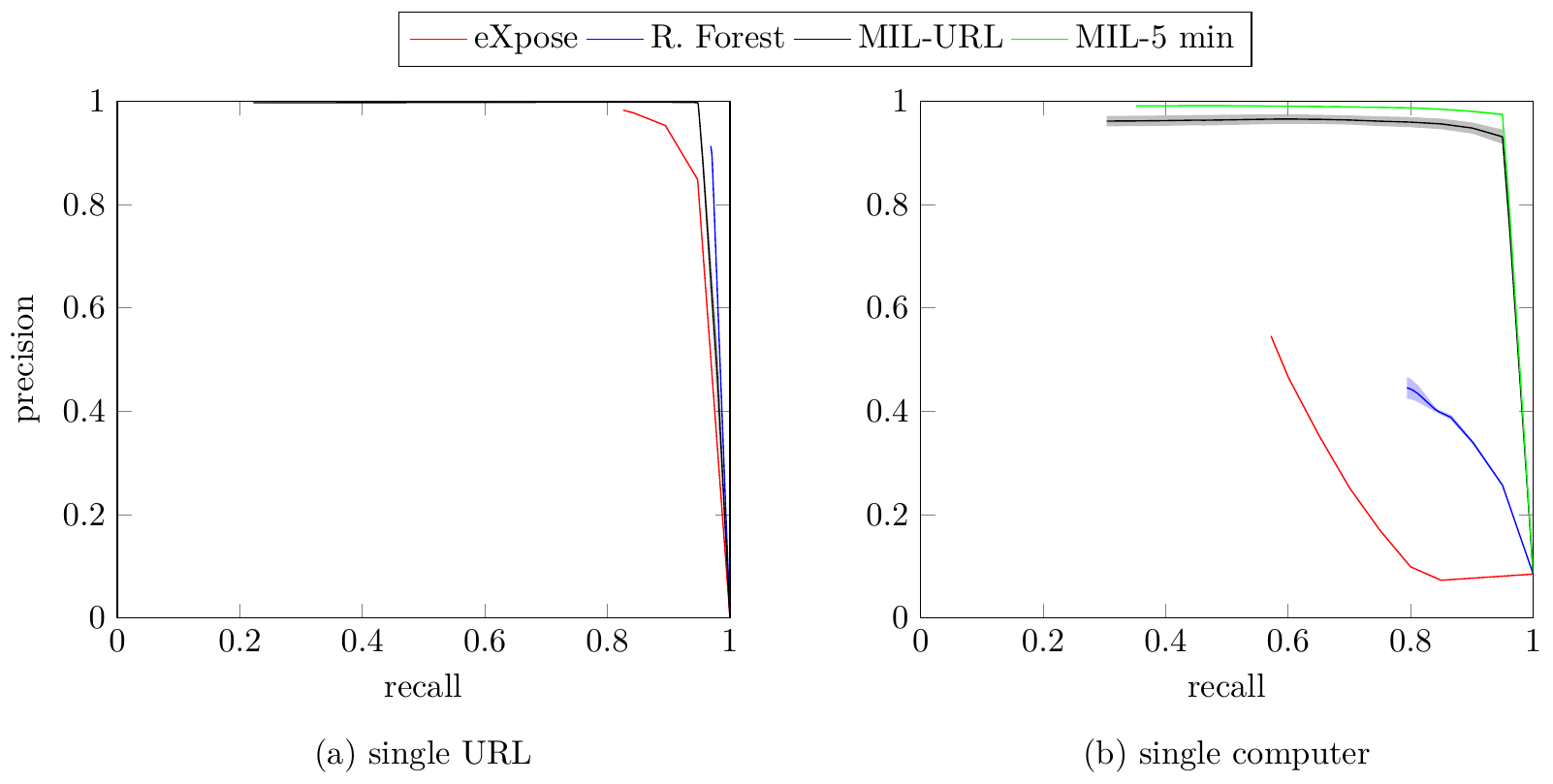}
    \caption{\label{fig:comparison}Precision-recall curves of compared detectors on data from 3rd November 2017. Graphs in the left plot are calculated from microscopic labels on individual URL strings whereas those in the right plot on macroscopic labels on individual computers.}
\end{figure*}

\section{Experimental results}
\label{sec:experimental}
This section presents the experimental comparison of the proposed classifiers MIL-URL and MIL-5min with the prior art based on human-designed features and Random Forests~\cite{machlica2017learning} (further called Random Forests) and the eXpose classifier based on convolutional neural networks (further called Convolution).

The classifiers are compared on four problems (or datasets). The first, called \emph{classification in the future}, corresponds to the case when the classifier is trained on data from the past (October 2017) and used on the data from the future ($3\mathrm{rd}$ November 2017). This comparison mimics the application in practice well. The second, called \emph{Grill test}, uses modified training as has been described above to estimate the accuracy on unseen or unlabelled data. The third, called \emph{CSIC}, is similar to the first case but performed on the publicly available CSIC dataset. The fourth problem uses the same scenario as the first one but measures the accuracy of classifiers in identifying the type of infection.

\subsection{Classification in the future}
Figure~\ref{fig:comparison}(a) shows precision-recall curves of classifiers of individual URLs (therefore the URL-5min classifier is missing). The proposed MIL-URL model offers only a slightly lower recall than the detector based on Random Forests, but with markedly higher precision. EXpose based on convolution neural networks is inferior to both R. Forest and MIL-URL. Figure~\ref{fig:comparison}(b) shows precision-recall curves of the same classifiers and also of the MIL-5min when one sample corresponds to one computer --- a scenario important for practitioners. We observe that both MIL-5min and MIL-URL dominate the prior art across all precision-recall space. Both classifiers keep precision above 95 percent with a similar recall. Prior art solutions have precision below 50 percent with worse recall.

The version of MIL-5min seems to be slightly better than the MIL-URL classifier. This should not be surprising as MIL-5min has more information about the infected computer on which it can base its decision, as is suggested by Ker's laws~\cite{ker2006batch,ker2017thesquare}. The superiority of MIL-5min is also important for label acquisition, as to train this classifier, it is sufficient to have labels on the level of 5-minute windows. These labels are simpler to obtain and therefore more precise than on the level of individual URLs.

The fact that Random Forests achieve good precision in classifying individual URLs but worse performance when the result is aggregated on the level of computers suggests that they suffer from higher false-positive rates.\footnote{While the loss in precision might seem to be puzzling, it is caused by the presence of computers infected by a noisy but easily detectable malware. When the level of sample granularity moves to a single computer, this large number of malicious URLs (samples) collapses to one, but a single false positive from a clean computer is still viewed as a single false positive.} To conclude, the proposed solutions based on multi-instance learning seems to be more precise in metrics better simulating practical applications.

\subsection{Dependency on training set size}
\begin{figure*}[t]
    \begin{subfigure}{0.33\textwidth}
        \includegraphics[width=0.9\linewidth]{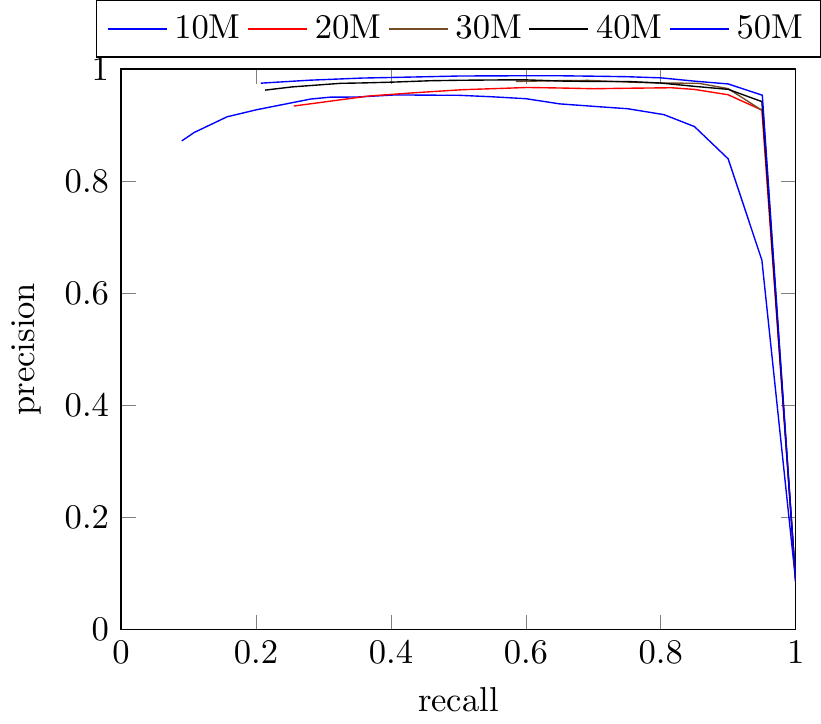}
        \caption{MIL-URL}
    \end{subfigure}
    \begin{subfigure}{0.33\textwidth}
        \includegraphics[width=0.9\linewidth]{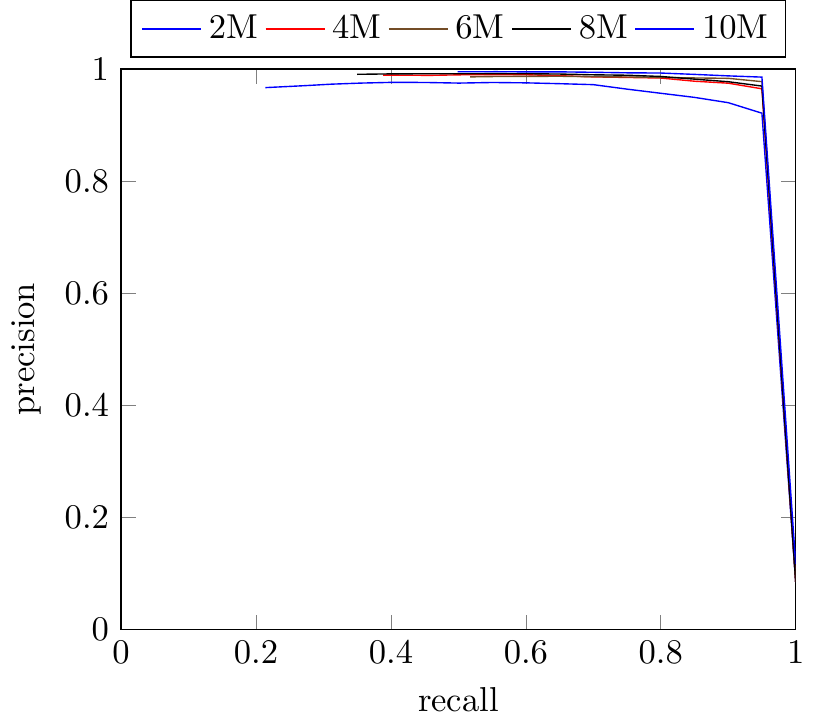}
        \caption{MIL-5min}
    \end{subfigure}
    \begin{subfigure}{0.33\textwidth}
        \includegraphics[width=0.9\linewidth]{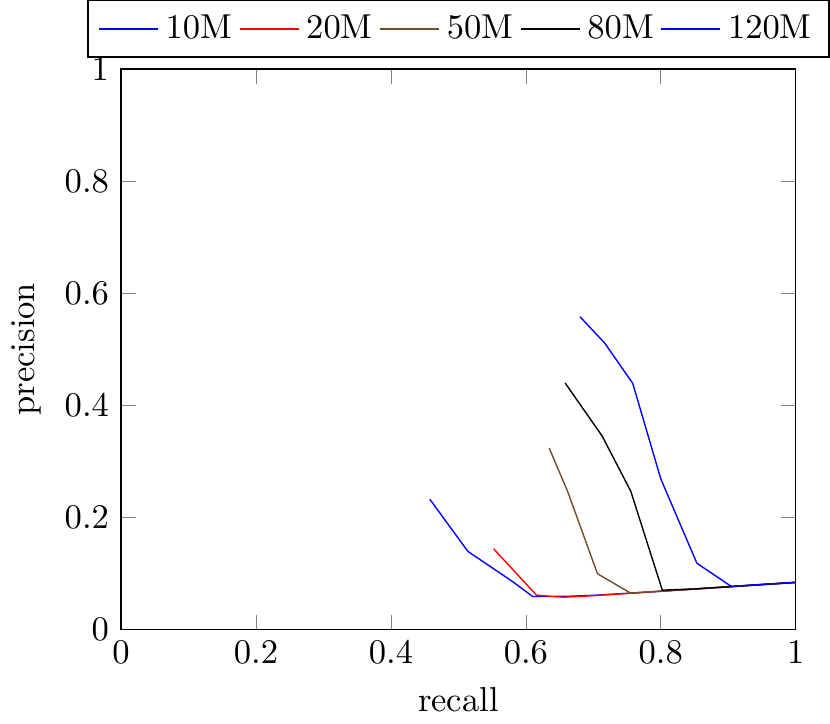}
        \caption{Random forests}
    \end{subfigure}
    \caption{\label{fig:trainingsamples} Precision-recall curves of the MIL-URL, MIL-5min classifiers and Random Forests for different number of positively labelled training samples. Since the MIL classifier had a fixed minibatch size of $1000$ samples (MIL-5min) and $5000$ samples (MIL-URL), the number of training samples was controlled by the number of training steps.}
\end{figure*}

Models based on neural networks, particularly deep ones, have a reputation of requiring a large number of samples. This section demonstrates that in the problem of URL classification, the Random Forest classifier requires even more. In fact, R. Forest classifier used in above section (Figure~\ref{fig:comparison}) used $2.4\times$ more samples than MIL-URL classifier. To study how the accuracy of different classifiers depends on the number of samples, we have varied (i) the number of training steps for the MIL classifiers and (ii) the number of samples per tree in the R. Forest classifiers, as these two hyper-parameters control the number of samples used in the training of corresponding classifiers. All other experimental settings were kept the same as above. Note though that in MIL-URL and in Random Forests one sample corresponds to one URL string, whereas in MIL-5min one sample corresponds to all URL strings collected for 5 minutes. Figure~\ref{fig:trainingsamples} shows PR curves of classifiers. We observe that the MIL-5min features the highest stability, as model trained on 4 million samples has almost the same accuracy as the one trained on 10 million samples used in Figure~\ref{fig:comparison}. Contrary, the R. Forest classifier features the lowest stability, as it continues to significantly improve even when it uses more than (120) samples ($12\times$ more than MIL-5min and $2.4\times$ more than MIL-URL model). The eXpose classifier was omitted from this study due to their poor performance and very high computational complexity (see Section~\ref{subsec:complexity} below).

\subsection{Generalization to unseen malware and new domains}
\begin{figure*}[ht]
    \centering
    \includegraphics[width=\textwidth]{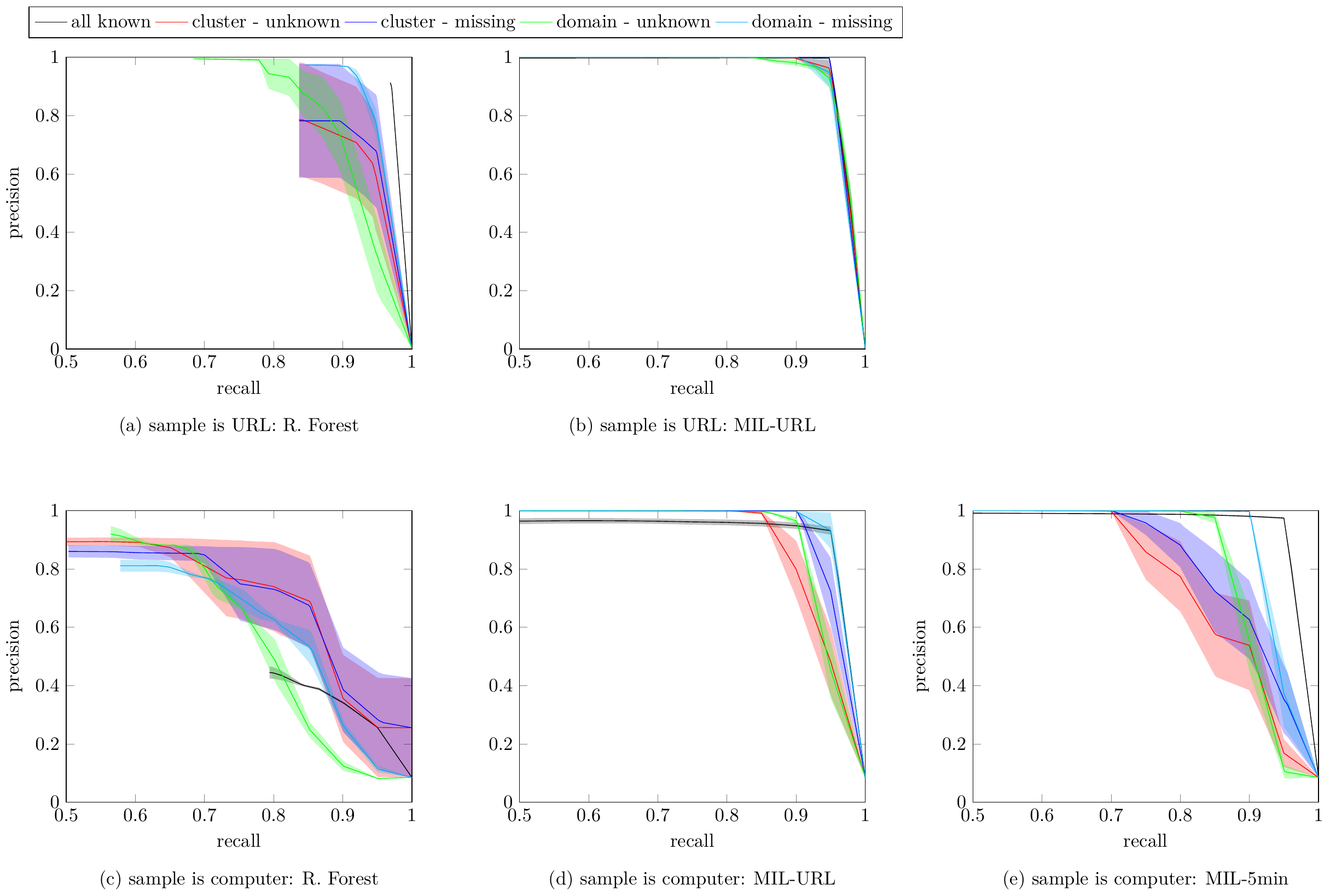}
    \caption{\label{url:grill-test}Precision-recall curves of MIL- 5min, Random Forest, and MIL-URL classifiers subjected to Grill-test on 3rd November 2017. All precision-recall curves assume a sample is a user to allow their comparison.}
\end{figure*}
In this section, the quality of detection of new hostnames/malware families of classifiers is, which shows how classifiers can generalize outside the training set. Again, due to poor performance and high computational cost, eXpose was omitted from this experiment.\footnote{One experiment requires 5 classifiers and the experiment was repeated 10 times.}

PR curves of the Grill-test are shown in Figure~\ref{url:grill-test}, wherein the top row a sample corresponds to a single URL and in the bottom row a sample corresponds to a single computer. We observe that in both cases, MIL models feature better generalization than R. Forest models. In almost all scenarios, the behavior is as expected, where classifiers trained on all known samples are better than classifiers with missing / mislabelled samples. The exception is the R. Forest classifier evaluated on the level of computers, where classifiers trained on data where some malware families were missing or mislabelled, the accuracy has improved. We believe that this is a result of overfitting to some malware families prevalent in the training set, yet rare in the testing set of 3rd November. Contrary to our expectations, the model classifying 5min windows is less robust than the model classifying individual URLs, which we cannot explain at the moment, but it can be caused by a smaller number of available training samples.

\subsection{Multi-class}
\begin{table}
    \centering
    \begin{tabular}{lrrr}
        \toprule
        Types                & Conv.         & R. Forest  & MIL-5min \\
        \midrule
        clean                & 99.7          & \textbf{99.9} & \textbf{99.9} \\
        trojan               & 34.0          & \textbf{99.9} & 95.1 \\
        ad injector          & \textbf{88.2} & 41.2          & 85.2 \\
        ransomware           & 0.6           & 57.1          & \textbf{100.0} \\
        malicious            & 6.5           & 4.4           & \textbf{84.4} \\
        PUA                  & 92.4          & 73.5          & \textbf{99.5} \\
        malware              & 23.3          & 92.6          & \textbf{99.1} \\
        information stealer  & 72.9          & 72.5          & \textbf{90.9} \\
        mal. cont. dist.     & 8.7           & 4.7           & \textbf{71.5} \\
        scareware            & 73.1          & 99.6          & \textbf{99.9} \\
        money scam           & 0.1           & 25.0          & \textbf{41.7} \\
        anon. software       & 15.1          & 35.6          & \textbf{90.0} \\
        banking trojan       & 32.4          & 7.1           & \textbf{96.7} \\
        spam tracking        & 0.3           & 10.1          & \textbf{11.9} \\
        click fraud          & 1.1           & \textbf{91.6} & 87.8 \\
        cryptocurrency miner & 55.1          & 99.0          & \textbf{99.8} \\
        \bottomrule
    \end{tabular}
    \caption{\label{tab:confusion} Accuracy of identification of the type of infection of the Convolution, Random Forest, and MIL-5min classifiers. The accuracy is measured on the level of computers from the traffic collected on 3rd November 2017. The best classifier result for a given type of malware is in bold.}
\end{table}

Table~\ref{tab:confusion} shows the accuracy of eXpose, R. Forest, and MIL-5min classifiers in identifying the type of infection of the computer. In almost all cases (except Trojan, ad-injector, and click fraud), the MIL-5min classifier is better than the prior art, sometimes very significantly. In the cases where the MIL approach is worse, it lags behind the best by less than $5\%.$

\subsection{CSIC dataset}
\label{subsec:exp:csic}
\begin{figure}
    \centering
    \includegraphics[width=0.9\columnwidth]{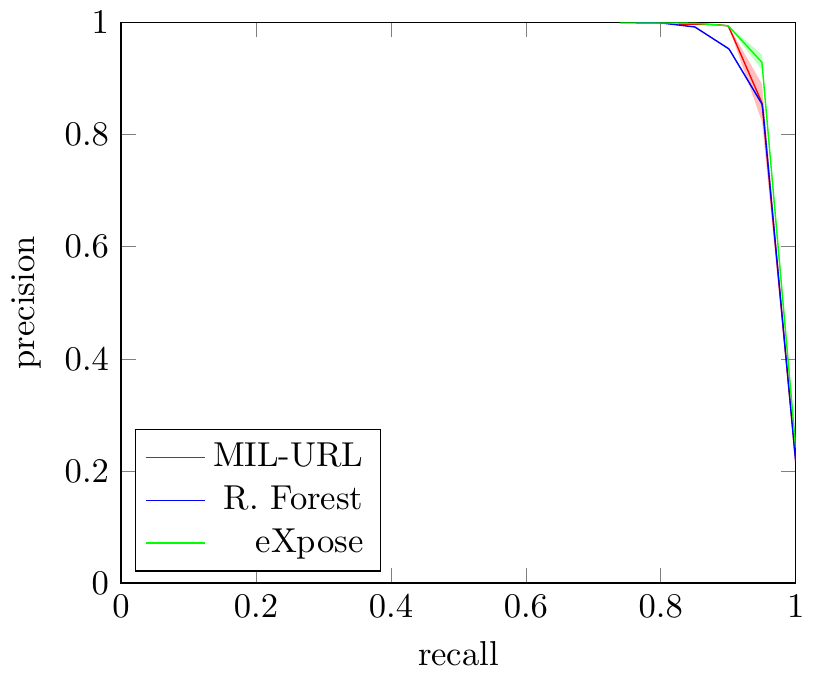}
    \caption{\label{fig:csic} Precision-recall curves of the MIL-URL, eXpose, and R. Forest classifiers on the CSIC dataset. 
    }
\end{figure}

This section compares all of the three evaluated approaches on the publicly available CSIC dataset of HTTP requests~\cite{perez2010applying}. Since the dataset is rather small and it contains only binary labels and all HTTP requests target the same host (\texttt{http://localhost:8080/}), the architectures of the classifiers were mildly altered to reflect this.

Specifically, the MIL-URL model has modeled only the path and query parts, as there is no reason to model the hostname, which is constant. The number of neurons in each layer was decreased to 40 (from 80 used in the corporate dataset) and the number of training iterations was decreased to 10 000. Finally, the cross-entropy has classified into just two classes (benign/malicious).

In eXpose, the hostname was removed from the URL string, since it is constant across the dataset and since the maximum length of URL is limited to 200 characters, keeping the hostname would decrease the expressive power of the model. Similarly to MIL-URL, the number of training iterations was decreased to 10 000.

Random Forest classifier was left almost intact with the exception that the training set for each tree was a random subset of 80\% of the training set. The rationale behind this was to increase the diversity of the ensemble to improve robustness.

Figure~\ref{fig:csic} shows the average Precision-Recall curves of 10 repetitions, where all available data were randomly split into a training set containing 80\% of samples and a testing set containing 20\% of samples. The sampling was stratified, which means that the class ratio was preserved. On this small problem, eXpose was mildly better in terms of recall than the MIL classifier and the R. Forest classifier was the worst. We believe that the superiority of eXpose is due to the simplicity of this problem, specifically because the number of possible attacks was much lower and attacks had very distinct signatures and the problem is smaller by nine orders of magnitude than previous problems in corporate networks.

\subsection{Extraction of IOCs}
\label{subsec:exp:IOC}
\begin{figure}
    \centering
    \includegraphics[width=0.9\columnwidth]{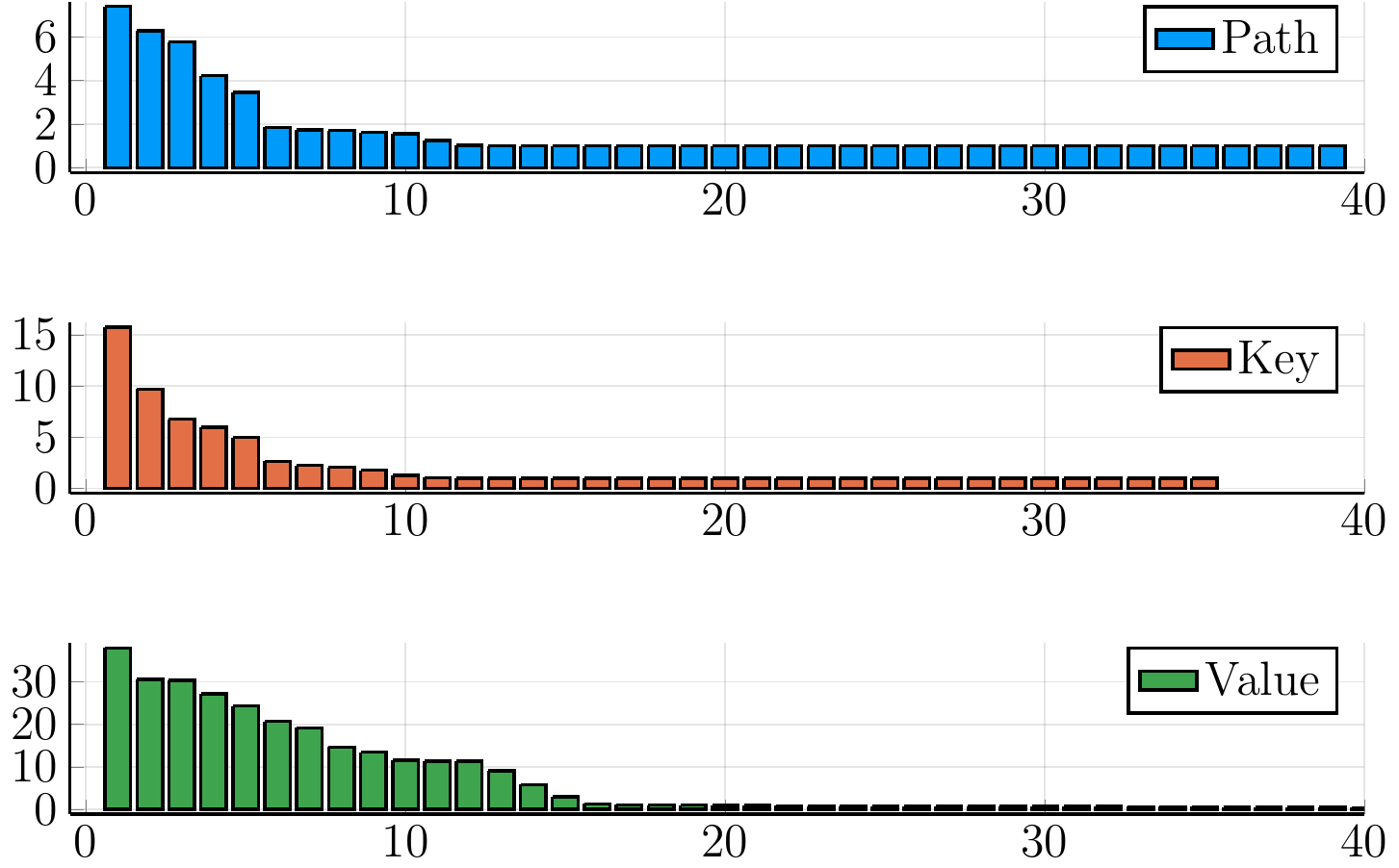}
    \caption{\label{fig:IOC} Ratios of peak values of neurons of the last layer before URL part aggregation for anomalous and normal data.}
\end{figure}
Indicators of compromise were extracted as described in Section~\ref{subsec:IOC} on the level of tokens in the path and in the query. The extraction on the level of traffic to domains was shown in~\cite{pevny2016discriminative} and we cannot repeat it here due to privacy restrictions on the Corporate dataset. Figure~\ref{fig:IOC}, therefore, shows the score, i.e. ratio of the output of neurons on anomalous and normal URL strings. According to these values, path and values seem to be more important for the detection of malicious URL strings than keys in the query string.

For path, the token with the overall highest output was \texttt{6909030637832563290.jsp.old}, for key \texttt{cantidadA} and value {\smaller \texttt{Espriella+Morcossessionid\%\allowbreak 3D12312312\%\allowbreak 26+\allowbreak username\%\allowbreak 3D\%\allowbreak 253C\%\allowbreak 2573\%\allowbreak 2563\%\allowbreak 2572\%\allowbreak 2569\%\allowbreak 2570\%\allowbreak 2574\%\allowbreak 253E\%\allowbreak 2564\%\allowbreak 256F\%\allowbreak 2563\%\allowbreak 2575\%\allowbreak 256D\%\allowbreak 2565\%\allowbreak 256E\%\allowbreak 2574\%\allowbreak 252E\%\allowbreak 256C\%\allowbreak 256F\%\allowbreak 2563\%\allowbreak 2561\%\allowbreak 2574\%\allowbreak 2569\%\allowbreak 256F\%\allowbreak 256E\%\allowbreak 253D\%\allowbreak 2527\%\allowbreak 2568\%\allowbreak 2574\%\allowbreak 2574\%\allowbreak 2570\%\allowbreak 253A\%\allowbreak 252F\%\allowbreak 252F\%\allowbreak 2561\%\allowbreak 2574\%\allowbreak 2574\%\allowbreak 2561\%\allowbreak 2563\%\allowbreak 256B\%\allowbreak 2565\%\allowbreak 2572\%\allowbreak 2568\%\allowbreak 256F\%\allowbreak 2573\%\allowbreak 2574\%\allowbreak 252E\%\allowbreak 2565\%\allowbreak 2578\%\allowbreak 2561\%\allowbreak 256D\%\allowbreak 2570\%\allowbreak 256C\%\allowbreak 2565\%\allowbreak 252F\%\allowbreak 2563\%\allowbreak 2567\%\allowbreak 2569\%\allowbreak 252D\%\allowbreak 2562\%\allowbreak 2569\%\allowbreak 256E\%\allowbreak 252F\%\allowbreak 2563\%\allowbreak 256F\%\allowbreak 256F\%\allowbreak 256B\%\allowbreak 2569\%\allowbreak 2565\%\allowbreak 2573\%\allowbreak 2574\%\allowbreak 2565\%\allowbreak 2561\%\allowbreak 256C\%\allowbreak 252E\%\allowbreak 2563\%\allowbreak 2567\%\allowbreak 2569\%\allowbreak 253F\%\allowbreak 2527\%\allowbreak 252B\%\allowbreak 2564\%\allowbreak 256F\%\allowbreak 2563\%\allowbreak 2575\%\allowbreak 256D\%\allowbreak 2565\%\allowbreak 256E\%\allowbreak 2574\%\allowbreak 252E\%\allowbreak 2563\%\allowbreak 256F\%\allowbreak 256F\%\allowbreak 256B\%\allowbreak 2569\%\allowbreak 2565\%\allowbreak 253C\%\allowbreak 252F\%\allowbreak 2573+\%\allowbreak 2563\%\allowbreak 2572\%\allowbreak 2569\%\allowbreak 2570\%\allowbreak 2574\%\allowbreak 253E\%\allowbreak 3F}}.
Although we do not know the precise meaning of the key ("cantidad" is in spanish "amount, count, number"), the value is a doubly-escaped, which after decoding (twice) using the service \url{https://www.motobit.com/util/url-decoder.asp} reads \texttt{=12312312\&username=<script>document.\allowbreak  location='http://attackerhost.example/\allowbreak cgi-bin/cookiesteal.cgi?'+document.cookie\allowbreak</script>?} suggesting a likely cookie stealing through cross-site scripting.

\subsection{Explaining the decision}
\label{subsec:exp:explanation}
\begin{table}
    \centering
    \begin{tabular}{rlr}
        \toprule
        Type  & Token                                  & Value \\
        \midrule
        path  & \texttt{"6909030637832563290.jsp.old"} & \( 0.999995 \) \\
        path  & \texttt{"tienda1"}                     & \( 1.22634 \times 10^{-9} \) \\
        path  & \texttt{"zarauz.jpg"}                  & \( 4.33613 \times 10^{-11} \) \\
        path  & \texttt{"miembros"}                    & \( -7.82587 \times 10^{-11} \) \\
        path  & \texttt{"imagenes"}                    & \( -0.857101 \) \\
        \bottomrule
    \end{tabular}
    \caption{\label{tab:explanation} The contributions of the individual tokens to the model prediction. Higher values mean more important tokens.}
\end{table}

The explanation algorithm described in Section~\ref{subsec:explanation} has been used on the CSIC dataset on tokens of a filename and keys and values. Again, the privacy restrictions on the Corporate dataset prevent us from using it in this test. As an example, the explainer is used on the positively classified URL string {\smaller \texttt{http://localhost:8080/\allowbreak tienda1/\allowbreak miembros/\allowbreak imagenes/\allowbreak zarauz.jpg/\allowbreak 6909030637832563290.jsp.old}}. The contributions of the individual tokens to the prediction of the model are listed in Table~\ref{tab:explanation}. As can be seen, the most important token is \texttt{6909030637832563290.jsp.old}, which nicely corresponds to indicators of compromise extracted in the previous section. 


\subsection{Computational complexity}
\label{subsec:complexity}
This section compares the complexity of all four models, namely a time to train the model, to classify all URLs in traffic of 5-minute windows of 1000 infected users (81090 URLs in total), and also the complexity of the model, measured by a size~\cite{risannen1978modeling} of a serialized Julia. Since the serialized model contains the data and necessary structures defining the model, we believe it to be a good proxy for the size used in~\cite{risannen1978modeling}.

The training and classification times of the MIL and Random Forest classifiers were measured on a single m5d.4xlarge AWS instance (64Gb of memory, 16 virtual Intel Xeon Platinum 8175M CPUs), the times of Convolution NN were measured on a single p2.xlarge AWS instance (60Gb of memory, 4 virtual Intel Xeon CPU E5-2686 CPUs with a single Tesla K80 GPU). Times are measured end2end, which means that they include all preprocessing with feature extraction in the case of Random Forest included.
\begin{table}
    \centering
    \begin{tabular}{lrrr}
        \toprule
                        & Training time  & Classification time & Model size\\
        \midrule
        eXpose          & 11 500         & 107.3               & 12M \\ 
        R. Forest       & 126 100        & 145.8               & 7.9M \\ 
        MIL-5min          & 34 000          & 6.9                 & 2.7M \\ 
        MIL-URL           & 1 100          & 6.4                 & 2.8M \\ 
        \bottomrule
    \end{tabular}
    \caption{\label{tab:complexity} Training and classification times (in seconds) of models used in experiments of this paper. The size of the minibatches of MIL-URL is $2\times5000$ samples (URLs), that of MIL-5min is $2\times1000$ samples (5 minute windows of users traffic, each of which is on average 14 URLs), and finally that of Convolution NN is $2\times 256$ samples. The size of the training set of Random Forest is $2\times 1.2 \cdot 10^{8}$ samples. The classification dataset contained the traffic of 1000 infected computers, which is approximately 81090 URLs.}
\end{table}

The measurements are shown in Table~\ref{tab:complexity}. We can see that the solution based on eXpose is the fastest to train, which is caused by the small size of the mini-batches mandated by limited GPU memory and also by employing a GPU, which for this type of task easily leads to 10 times faster training. On the other hand, eXpose is the largest model and the classification time is the second highest, caused by slow loading of data to the GPU. Random Forest was the most expensive to train and also the classification time is the highest, more than 20 times that of MIL and 1.5 times that of eXpose.

The training time of MIL-URL is the fastest, while that of MIL-5min is the second slowest. This discrepancy is caused by the size of the mini-batches of $1000$ 5-min windows of MIL-5min, which is approximately 50 000 URLs (opposed to 5000 URLs of the MIL-URL model). On the other hand, in both cases, the classification time is more than an order of magnitude faster than that of the prior art. Also, the model is the lowest complexity as measured by the stored serialized model size.

\section{Related work}
\label{sec:related}
The evolution of methods to detect malicious URLs follows the evolution of machine learning. Early, but still used methods~\cite{kruegel2003anomaly,zarras2014automated,machlica2017learning,li2018method} rely on human-designed features. For example~\cite{kruegel2003anomaly} uses the length of the URL, frequency of selected characters or occurrence of special tokens in the query part. For each type of features, it designs a method to detect anomalies, such as the chi-square test for the frequency of characters. A detector for spam, phishing and malware in~\cite{choi2011detecting} uses the number of, average length and maximum length of the domain, path, and query tokens together with spam, phishing and malware SLD hit ratio brand name presence. The most complete set of features known to us~\cite{machlica2017learning,li2018method} was used in the experimental section in the Random Forest classifier. The feature set contains (i) characterization of distinct patterns in the URL string, including the length of the URL, the vowel ratio, the consonant ratio, the number of special characters ('!', '-', '\_', ',', '@', '\#', '\%', '+', ':', ';'), the upper case ratio, the lower case ratio, and the proportion of the digits; (ii) common statistical features of domain names, including domain name levels, character type distribution ratio, and top-level domain name; (iii) the overall length of the path and the number of directories; (iv) and finally filename suffix and its length.

Refs.~\cite{ma2009beyond} and~\cite{rajalakshmi2018effective} avoid the design of features by either creating a dictionary of observed tokens or their 3--8grams. With that, they represent the URL by their one-hot encoded presence. These works build upon the progress of training linear models with a large dimension on very sparse samples.

Recent advances in neural networks from vision and language modeling were used in~\cite{saxe2017expose,lin2018character}, and~\cite{woodbridge2016predicting}, where the URL is treated as a sequence of characters truncated to 200 characters, as according to the references 95\% of URLs are shorter. CNN from~\cite{saxe2017expose} is used in the comparison. Unlike the proposed method it has to learn to parse the URL to utilize the structure.

The closest work to this is~\cite{zhang2017deeplearning} and our earlier work~\cite{dedic2017hierarchical}, which divides the URL into a hostname, path and query modeled separately either by a convolution over tokens embedded to Euclidean space using word2vec~\cite{mikolov2013distributed} or by the multiple-instance learning framework~\cite{pevny2017using}. The proposed model can be seen as an evolution that uses a more sophisticated model of query and models all traffic of the computer instead of a single URL (which gives the model ability to be trained from coarse labels and to detect infections merely changing distribution on otherwise legitimate servers).

In~\cite{ma2009identifying}, URL features are supplemented by information about the server, such as data from WHOIS, IP prefix, autonomous system number, geographic location, and connection speed. Although the presented model can be modified to use these features, they were avoided since we wanted to compare to the prior art on URLs only.

As some malware has a very specific header of the HTTP request, Ref.~\cite{zarras2014automated} creates a template for each family and measures the distance from it. Thus, it essentially builds a 1-nearest neighbor classifier with a custom distance. The proposed hierarchical models can include the model of HTTP header, yet the data were not available to us and as mentioned above, utilizing them would limit the comparison to the prior art.

So far, we have reviewed models detecting individual URL strings. Recognizing the deficiencies, in~\cite{bartos2016optimized} the subject of classification is a set of connections between the hostname and a particular server. The representation is inspired by models from computer vision, but it cannot be easily extended to the traffic to multiple servers. A model for the same subject is proposed in~\cite{kohout2018network}, but the scaling in higher dimensions is dubious. Moreover, unlike the presented model both methods rely on hand-crafted features.

The clustering of binaries executed in a sandbox based on their URLs is treated in~\cite{perdisci2010behavioral}. To calculate distance, it is proposed to use a weighted sum of Levenstein distance between strings, Jaccard index between parameters, etc. Since the goal of the proposed work is classification, the distance function of~\cite{perdisci2010behavioral} is not well suited for this problem, as k-NN or SVM classifiers do not scale.

Last, we mention our previous work~\cite{pevny2016discriminative}, where the network host is modeled by a set of domains and each domain by a set of HTTP messages exchanged with it (see Section~\ref{sub:Traffic}). While~\cite{pevny2016discriminative} requires URL strings to be described by a set of features, whereas the proposed model extends it such that it requires the representation of individual string tokens only (for example by 3-grams).

\section{Conclusion}
\label{sec:conclusion}
The main goal of this work was to replicate in the field of network security 
the success of convolution neural networks in computer vision and other areas in removing
human-designed features. This has been achieved by removing time
and spatial dependencies and by nesting multiple instance learning problems.
The proposed framework can classify a set of all connections from a single
computer while relying on features describing only string tokens. To the best of our
knowledge, this is the first work of its type.

Experimental results have demonstrated that the proposed model outperforms, 
sometimes significantly, the prior art in the problem of identifying infected computers within a computer network and classifying the type of infection. 

We believe that the proposed approach will serve as a blueprint approach, how domains with complicated hierarchies can be elegantly handled by straightforward nesting of multiple-instance learning problems. We have
therefore released a library simplifying this task at \texttt{https://github.com/pevnak/Mill.jl}.
We also believe that there is a lot of space for further improvement, for example
by utilizing local dependencies by convolution.
\bibliographystyle{plain}

\end{document}